\documentclass[preprint,prd,superscriptaddress]{revtex4}

\usepackage{graphicx}
\usepackage{dcolumn}
\usepackage{bm}

\begin{document}

\preprint{RIKEN-TH-98}
\preprint{UTHEP-545}
\preprint{KEK-CP-194}
\preprint{YITP-07-27}
\preprint{NTUTH-07-505D}

\title{
Two-flavor lattice QCD
in the $\epsilon$-regime 
and chiral Random Matrix Theory
}

\newcommand{\RIKEN}{
  Theoretical Physics Laboratory, RIKEN,
  Wako 351-0198, Japan
}

\newcommand{\Taiwan}{
  Physics Department and Center for Theoretical Sciences, 
                National Taiwan University, Taipei, 10617, Taiwan  
}

\newcommand{\Tsukuba}{
  Graduate School of Pure and Applied Sciences, University of Tsukuba,
  Tsukuba, Ibaraki 305-8571, Japan
}
\newcommand{\BNL}{
  Riken BNL Research Center, Brookhaven National Laboratory, Upton,
  NY11973, USA
}
\newcommand{\KEK}{
  High Energy Accelerator Research Organization (KEK),
  Tsukuba 305-0801, Japan
}
\newcommand{\GUAS}{
  School of High Energy Accelerator Science,
  The Graduate University for Advanced Studies (Sokendai),
  Tsukuba 305-0801, Japan
}

\newcommand{\YITP}{
  Yukawa Institute for Theoretical Physics, 
  Kyoto University, Kyoto 606-8502, Japan
}

\newcommand{\HUDP}{
  Department of Physics, Hiroshima University,
  Higashi-Hiroshima 739-8526, Japan
}

\author{H.~Fukaya}
\affiliation{\RIKEN}

\author{S.~Aoki}
\affiliation{\Tsukuba}
\affiliation{\BNL}

\author{T.W.~Chiu}
\affiliation{\Taiwan}

\author{S.~Hashimoto}
\affiliation{\KEK}
\affiliation{\GUAS}

\author{T.~Kaneko}
\affiliation{\KEK}
\affiliation{\GUAS}

\author{H.~Matsufuru}
\affiliation{\KEK}

\author{J.~Noaki}
\affiliation{\KEK}

\author{K.~Ogawa}
\affiliation{\Taiwan}

\author{T.~Onogi}
\affiliation{\YITP}

\author{N.~Yamada}
\affiliation{\KEK}
\affiliation{\GUAS}

\collaboration{JLQCD collaboration and TWQCD collaboration}
\noaffiliation


\begin{abstract}
  The low-lying eigenvalue spectrum of the QCD Dirac operator in
  the $\epsilon$-regime is expected to match with that of 
  chiral Random Matrix Theory (ChRMT).
  We study this correspondence for the case including
  sea quarks by performing two-flavor QCD simulations on the
  lattice.
  Using the overlap fermion formulation, which preserves exact chiral symmetry 
  at finite lattice spacings, we push the sea quark mass down to $\sim$ 3~MeV
  on a $16^3\times 32$ lattice at a lattice spacing $a\simeq$ 0.11~fm.
  We compare the low-lying eigenvalue distributions and 
  find a good agreement with the analytical predictions of ChRMT.
  By matching the lowest-lying eigenvalue we extract the
  chiral condensate, 
  $\Sigma^{\overline{\mathrm{MS}}}(2\mbox{~GeV}) 
  = (251\pm 7\pm 11 \mbox{~MeV})^3$, 
  where errors represent statistical and higher
  order effects in the $\epsilon$ expansion.
  We also calculate the eigenvalue distributions on the
  lattices with heavier sea quarks at two lattice spacings. 
  Although the $\epsilon$ expansion is not applied for those
  sea quarks, we find a reasonable agreement of the Dirac
  operator spectrum with ChRMT.
  The value of $\Sigma$, after extrapolating to the chiral
  limit, is consistent with the estimate in the $\epsilon$-regime. 
\end{abstract}

\maketitle

\section{Introduction}
Numerical simulations of QCD on the lattice suffer from
various sources of systematic errors, such as finite lattice
spacing $a$, finite volume $V$, and larger quark masses
$m$ than those in the nature.
Each of these needs to be eliminated by an extrapolation
using several independent simulations.
In particular, the extrapolation in the quark mass to the
chiral (or physical) limit is non-trivial, because most
physical quantities have non-analytic dependence on the
quark masses due to pion loop effects as predicted by
chiral perturbation theory (ChPT). 
In order to reproduce such non-analytic behavior, the
physical volume must be increased as the chiral limit is
approached such that the pion Compton wavelength fits in the
box. 
Therefore, in practice the chiral extrapolation must be done
with a limited range of quark masses, which is
a potential source of large systematic uncertainty.
This becomes more problematic when the chiral symmetry
is explicitly violated by the fermion formulation on the
lattice, since the standard ChPT cannot be used as a guide
in the extrapolation and the chiral extrapolation must be
combined with the continuum extrapolation.

An alternative approach is to study the $\epsilon$-regime
of QCD 
\cite{Gasser:1987ah, Hansen:1990un,Hansen:1990yg,Leutwyler:1992yt}
on the lattice.
In this regime the quark mass is set close to the chiral
limit while keeping the physical volume finite.
The system suffers from a large finite volume effect, but it
can be systematically calculated by ChPT, because the
pion field dominates the low energy dynamics of the system
and the effects of other heavier hadrons become sub-dominant.
It means that the low energy constants appearing in  ChPT
Lagrangian can be extracted from the lattice calculation in
the $\epsilon$-regime by comparing with  ChPT
predictions. 
Since a small violation of chiral symmetry gives large
effects in the $\epsilon$-regime, the lattice fermion
formulation must fully respect the chiral symmetry.

The $\epsilon$-regime is reached by reducing the quark mass
$m$, at a finite volume $V=L^3T$, down to the region where
the pion mass $m_\pi$ satisfies the condition 
\begin{equation}
  \label{eq:e-regime}
  1/\Lambda_{\rm QCD}\ll L \ll 1/m_{\pi} ,
\end{equation}
where $\Lambda_{\rm QCD}$ denotes the QCD scale.
Under the condition (\ref{eq:e-regime}), the zero momentum
modes of the pion field give the dominant
contribution since the energy of finite momentum modes is
too large to excite.
In this way, ChPT is organized as an expansion in terms of
the parameter 
$\epsilon^2 \sim m_\pi/\Lambda_{\rm UV}\sim p^2/\Lambda^2_{\rm UV}$ 
where $\Lambda_{\rm UV}$ is the ultraviolet cut-off of ChPT
(typically taken to be $4\pi F_\pi$ with $F_\pi$ the pion
decay constant). 
Since the quantum correction of the zero-modes is not
suppressed in the $\epsilon$-regime and the path
integral over SU($N_f$) manifold must be explicitly carried out,
the partition function and other physical quantities show
remarkable sensitivity to the topology of the gauge field.

At the leading order of the $\epsilon$-expansion, 
the partition function of ChPT
is equivalent to that of 
 chiral Random Matrix Theory (ChRMT) 
\cite{Shuryak:1992pi,Smilga:1995nk,Verbaarschot:2000dy,%
  Damgaard:2000ah,Akemann:2006ru}
 at any fixed topological charge.
Moreover, from the symmetry of the Dirac operator,
the low-lying QCD Dirac spectrum is expected to be 
in the same universality class of ChRMT.
ChRMT thus provides a direct connection between Dirac eigenvalues
and the effective theory describing the dynamical
chiral symmetry breaking. 
One of the most convenient predictions of ChRMT is the
distribution of individual eigenvalue, which can be directly
compared with the lattice data.
Such comparison has been done mainly in the quenched
approximation 
\cite{Edwards:1999ra,Bietenholz:2003mi,Giusti:2003gf, Wennekers:2005wa},
except for a work using the reweighting technique
\cite{Ogawa:2005jn} or for some recent attempts of carrying
out dynamical fermion simulation on coarse lattices
\cite{DeGrand:2006nv,Lang:2006ab}.
The eigenvalue spectrum in those calculations shows a
    good agreement with the prediction of ChRMT as
far as the lattice volume is large enough 
$\gtrsim (1.5\mbox{~fm})^4$.

In this work we perform lattice QCD simulations in and
out of the $\epsilon$-regime including two light flavors of
dynamical quarks. 
Since we are interested in the consequences of chiral
symmetry breaking, we employ the Neuberger's overlap-Dirac
operator \cite{Neuberger:1997fp,Neuberger:1998wv},
which preserves exact chiral symmetry \cite{Luscher:1998pq}
at finite lattice spacings.
The exact chiral symmetry is also helpful for
numerical simulations in the $\epsilon$-regime, because
the lowest-lying eigenvalue of the Hermitian
overlap-Dirac operator is bounded from below (by a small but
finite mass term) and no numerical instability occurs.
The space-time volume of our lattice is 
$L^3\times T = 16^3\times 32$ 
with the lattice spacing $a\sim$ 0.11--0.125~fm.  
The gauge field topology is fixed to the trivial topological
sector by introducing the extra Wilson fermions and ghosts
\cite{Fukaya:2006vs}.
We perform the Hybrid Monte Carlo simulation with the sea
quark mass around 3~MeV, which corresponds to the
$\epsilon$-regime: the expected pion Compton wavelength is
comparable to the lattice extent $m_\pi L\simeq 1$.
The numerical cost for such a small sea quark mass is very
expensive in general, but it is not prohibitive on the small
lattice as required in the $\epsilon$-regime simulation. 
We also carry out simulations at several quark masses 
    roughly in the region $m_s/6$-$m_s$ with $m_s$ the physical
    strange quark mass, which are out of the $\epsilon$ regime.


We study the eigenvalue spectrum of the overlap-Dirac
operator on the configurations generated with these dynamical
quarks.
A good agreement of the low-lying eigenvalue spectrum with
 ChRMT predictions has already been reported in our
earlier paper \cite{Fukaya:2007fb} for the run in the
$\epsilon$-regime.
The present paper describes our analysis in more detail.
Since  ChRMT provides the distribution of individual
eigenvalues, the test of the agreement can be made using the
information on the shape of the distribution, not just using
the average values.
We find a good agreement of the lowest-lying eigenvalue
distribution by analyzing its several moments.
If we look at higher eigenvalues, the agreement becomes
marginal, because there are contaminations from the bulk of
the eigenvalue spectrum corresponding to finite momentum
pion states and other higher excited states, which are not
described by ChRMT.
We study the bulk eigenvalue spectrum and identify the
region where the analysis in the $\epsilon$-regime is
applied. 

A direct output from the comparison of the eigenvalue
spectrum is the value of chiral condensate $\Sigma$.
We extract $\Sigma$ from the lowest-lying eigenvalue in the
$\epsilon$-regime.
For comparison we also calculate it on heavier quark mass
lattices and extrapolate them to the chiral limit.
Although the leading order relations in the $\epsilon$
expansion is not valid for these lattices, the result in the
chiral limit shows remarkable agreement with the direct
calculation in the $\epsilon$-regime.
We convert the value of $\Sigma$ obtained on the lattice to
the common definition in the continuum renormalization
scheme $\overline{\mathrm{MS}}$ using the non-perturbative
renormalization (NPR) technique through the RI/MOM scheme
which is a 
    regularization independent scheme based on the 
    Green's functions of the offshell quark
\cite{Martinelli:1994ty}.

This paper is organized as follows.
In Section~\ref{sec:ChRMT}, we review  ChRMT calculations of
the Dirac eigenvalue spectrum. 
The details of the numerical simulations are described in
Section~\ref{sec:simulation}, and the results of the low-lying modes
in the $\epsilon$-regime is discussed in Section~\ref{sec:low-modes}.
The low-mode spectrum in the $p$-regime 
are presented in Section~\ref{sec:p-regime}.
In Section~\ref{sec:bulk} we also study the higher
eigenvalue spectrum.
Our conclusions are given in Section~\ref{sec:conclusion}.

\section{Chiral Random Matrix Theory}
\label{sec:ChRMT}

In the $\epsilon$-regime the low-lying eigenvalue spectrum
of $N_f$-flavor QCD Dirac operator matches with that of
Chiral Random Matrix Theory (ChRMT) 
\cite{Shuryak:1992pi,Smilga:1995nk,Verbaarschot:2000dy,Damgaard:2000ah}
up to a scale factor as described below.
This can be derived by identifying the partition function of
ChRMT
\begin{equation}
  \label{eq:RMT}
  Z_Q(\hat{m})=\int dW
  e^{-\frac{N}{2}\mbox{tr}W^{\dagger}W}\det 
  \left(
    \begin{array}{cc}
      \hat{m} & W \\ -W^{\dagger} & \hat{m}
    \end{array}
  \right)^{N_f},
\end{equation}
with the QCD partition function in the $\epsilon$-regime.
Since the dependence on the global topology becomes manifest
in the $\epsilon$-regime, we work in a fixed topological
sector $Q$.
Here, $W$ is a complex $(n+Q)\times n$ matrix, and $N\equiv 2n+Q$.
The parameter $\hat{m}$ plays a role of quark mass.
In the limit of large $N$, the partition function
(\ref{eq:RMT}) can be modified to the form describing the
zero-momentum mode of ChPT \cite{Shuryak:1992pi}
\begin{equation}
  Z_Q(\hat{m}) = \int_{U\in U(N_f)} DU (\det U)^Q 
  \exp\left[ \frac{N}{2}\,\mathrm{tr}(\hat{m}U+\hat{m}U^{\dagger})
    + O(\hat{m}^2)\right],
\end{equation}
from which one can identify $N \hat{m}= m\Sigma V$.

The advantage of ChRMT (\ref{eq:RMT}) is that the eigenvalue
distribution of the matrix $W^\dagger W$ is analytically
known \cite{Damgaard:2000ah}.
Here we reproduce the known result for the case of two
degenerate flavors and zero topological charge, which is
relevant in this work.

Let us consider the $k$-th lowest microscopic eigenvalue 
$\zeta_k = N x_k$, with $x_k$ the $k$-th eigenvalue of 
$\sqrt{W^\dagger W}$.
The distribution of $\zeta_k$ is written as
\begin{equation}
  {p}_{k}(\zeta_k;\mu)=
  \int_0^{\zeta_k} d\zeta_1 \int_{\zeta_{1}}^{\zeta_k} d\zeta_2 
  \cdots \int_{\zeta_{k-2}}^{\zeta_k} d\zeta_{k-1}
  \omega_{k}(\zeta_1 , \cdots , \zeta_k;\mu ),
\end{equation}
where $\mu\equiv N\hat{m} = m\Sigma V$.
The form of $\omega_k(\zeta_1,\cdots,\zeta_k;\mu)$ is
analytically known in the microscopic limit, {\it i.e.}
$n\to \infty$ while $\mu$ is kept fixed:
\begin{eqnarray}
   \omega_k(\zeta_1,\ldots,\zeta_k; \mu)=
   \mathrm{const.}\,
   e^{-\zeta_k^2/4} 
   (\prod_{i=1}^k \zeta_i)
 \frac{
   [\prod_{j=1}^{k-1} (\zeta_k^2-\zeta^2_j)^2] (\zeta_k^2+\mu^2)^2
 }{
   \prod_{i>j}^{k-1} (\zeta_i^2-\zeta^2_j)^2 
   \prod_{j=1}^{k-1} (\zeta_j^2+\mu^2)^2
 }
 \frac{\det[B]}{\det[A]}.
\end{eqnarray}
The matrices $A$ and $B$ are given by 
\begin{equation}
  A=\left(
    \begin{array}{cc}
      I_0(\mu) & \mu^{-1}I_1(\mu)\\
      \mu I_1(\mu) & I_0(\mu)
    \end{array}
  \right),
  \;
  B_{ij}=
  \left\{
    \begin{array}{cc}
      \tilde{\mu}^{j-3} I_{j-3}(\tilde{\mu}) & (i=1)\\
      \tilde{\mu}^{j-4} I_{j-4}(\tilde{\mu}) & (i=2)\\
      \tilde{\zeta}_i^{j-3} I_{j-3}(\tilde{\zeta}_i) &
      (3\leq i \leq k+1) \\
      \tilde{\zeta}_i^{j-4} I_{j-4}(\tilde{\zeta}_i) &
      (k+2 \leq i \leq 2k)
    \end{array}
  \right. 
  \;
  (1\leq j \leq 2k),
\end{equation}
where $\tilde{\zeta}_i\equiv \sqrt{\zeta_k^2-\zeta_i^2}$
and $\tilde{\mu}\equiv \sqrt{\zeta_k^2+\mu^2}$.
$I_i(x)$'s are the modified Bessel functions.

\begin{figure}[tbp]
  \centering
  \includegraphics[width=10cm,clip=true]{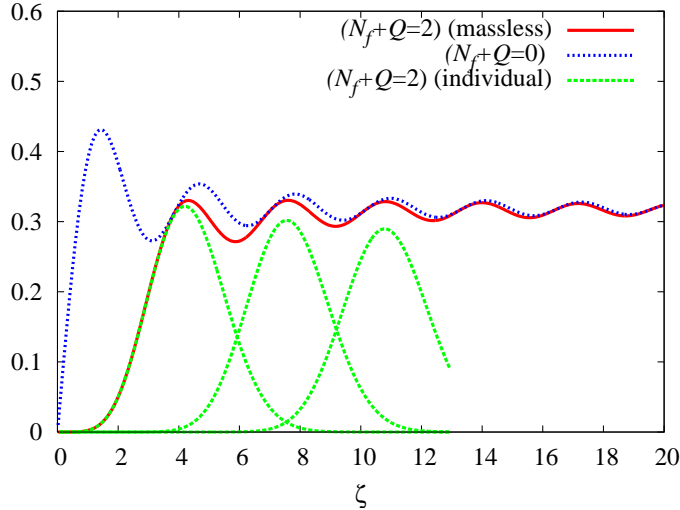}
  \caption{
    Low-lying spectral density in the massless limit
    $\rho_{\mbox{\tiny RMT}}(\zeta;0)$ (solid curve) and its decomposition to
    individual eigenvalues $p_k(\zeta_k;0)$ 
    (dashed curves, for $k$ = 1, 2 and 3). 
    The dotted curve represents the distribution in the
    infinite sea quark mass limit $\rho_{\mbox{\tiny RMT}}(\zeta;\infty)$,
    which corresponds to the quenched theory.
  }
  \label{fig:dist}
\end{figure}

The spectral density is given by a sum of the individual distributions
\begin{equation}
\label{eq:density}
  \rho_{\mbox{\tiny RMT}}(\zeta;\mu) \equiv \sum_k p_k(\zeta;\mu).
\end{equation}
In the massless and the infinite mass (or quenched) limit,
it can be written in a simple form,
\begin{eqnarray}
  \label{eq:density2}
  \rho_{\mbox{\tiny RMT}}(\zeta;0)&=& 
  \frac{\zeta}{2}
  \left(J_2^2(\zeta)-J_3(\zeta)J_1(\zeta)\right),\nonumber\\
  \rho_{\mbox{\tiny RMT}}(\zeta;\infty)&=& 
  \frac{\zeta}{2}
  \left(J_0^2(\zeta)+J^2_1(\zeta)\right),
\end{eqnarray}
where $J_i(\zeta)$ denotes the Bessel functions of 
the first kind.
Their shape and the individual eigenvalue distributions
are shown in Figure~\ref{fig:dist}.

\begin{figure}[tbp]
  \centering
  \includegraphics[width=10cm,clip=true]{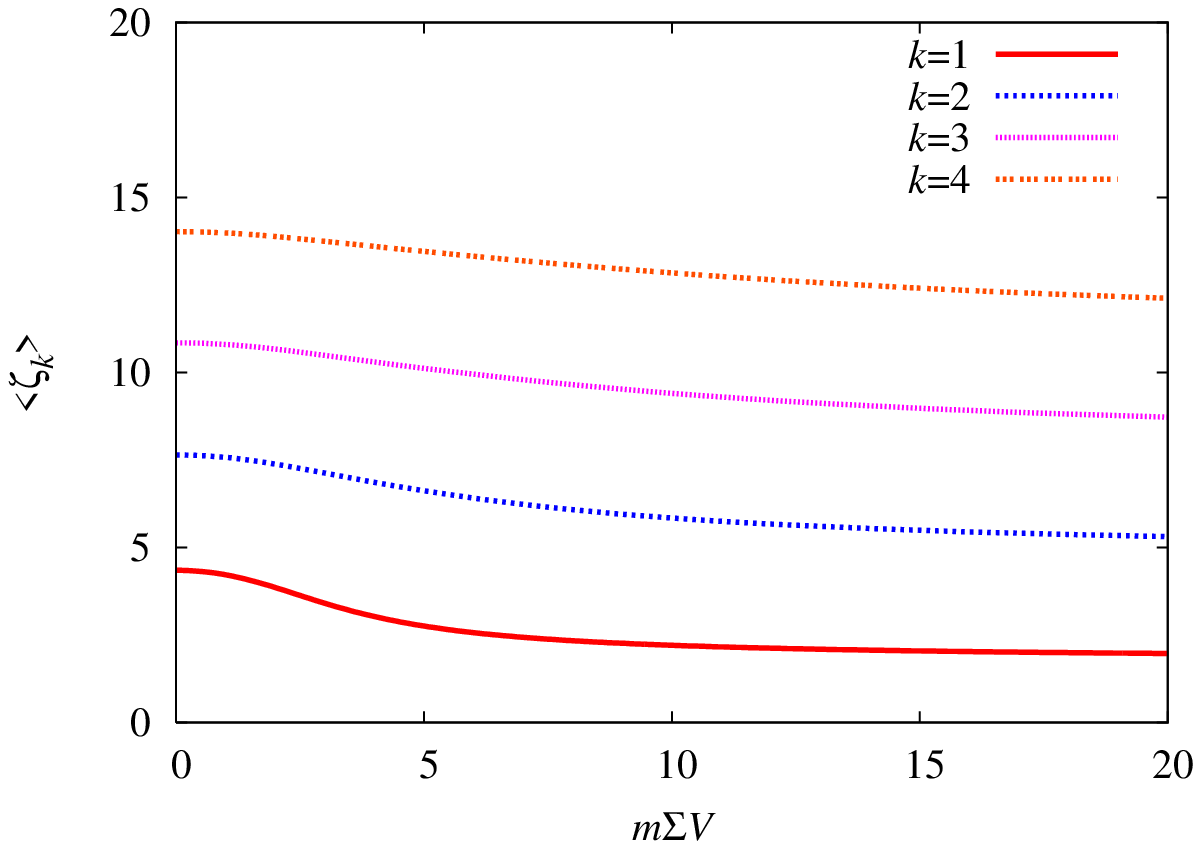}\\
  \includegraphics[width=10cm,clip=true]{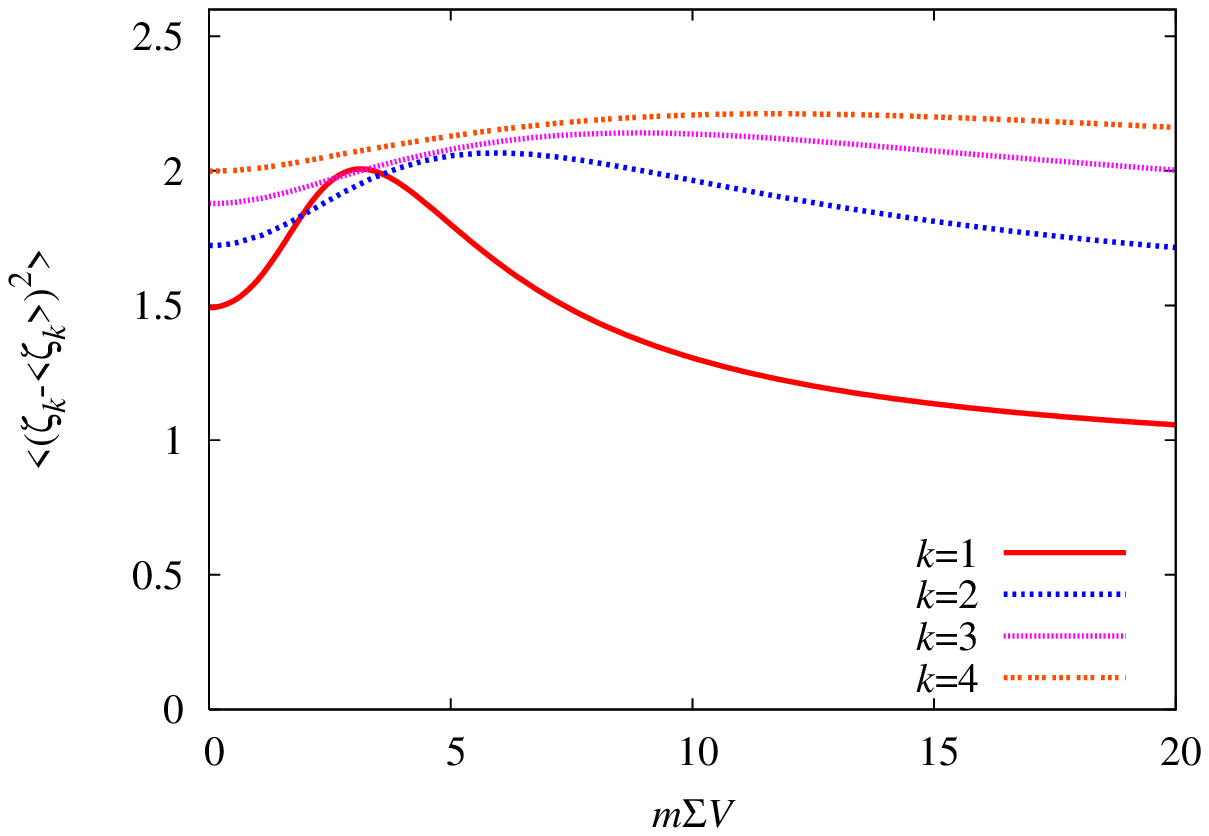}
  \includegraphics[width=10cm,clip=true]{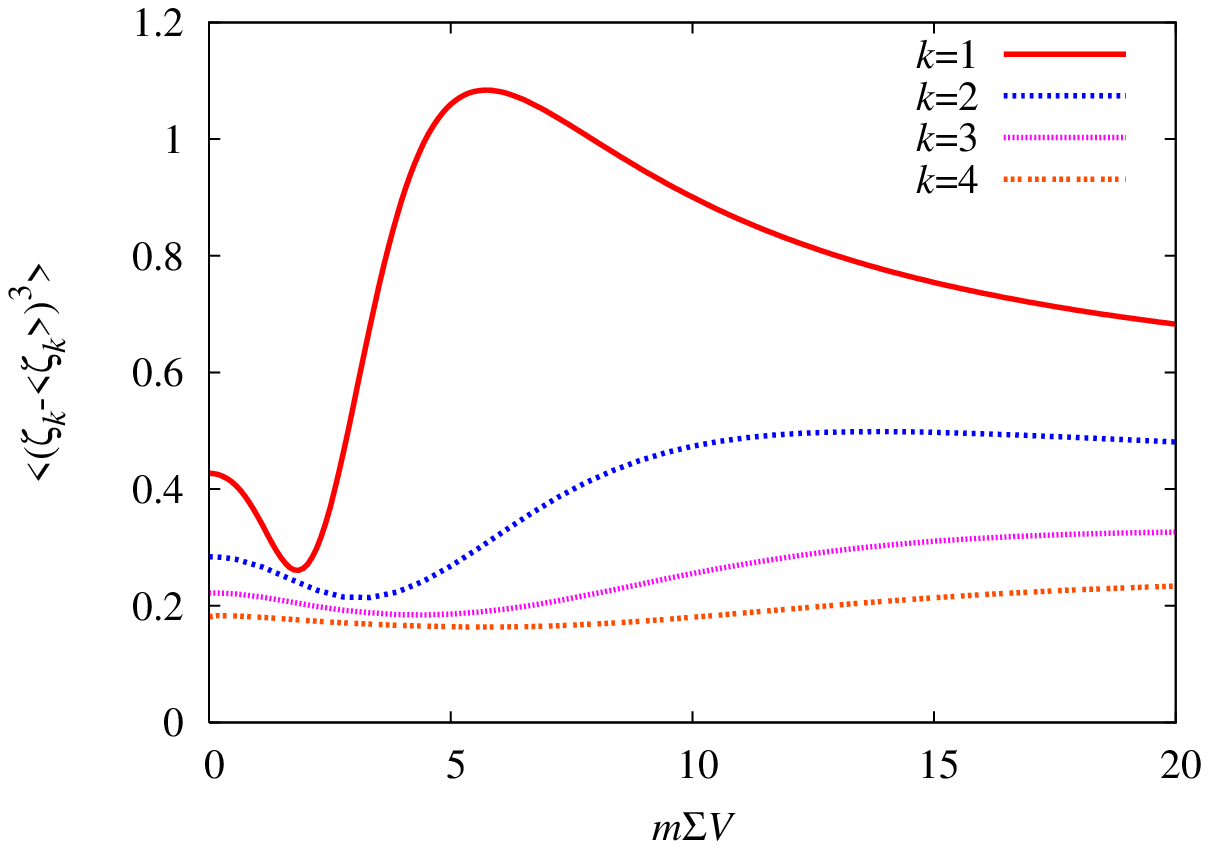}\\
  \caption{
    First (top), second (middle) and third (bottom)
    moments of the lowest-lying eigenvalues ($k$ = 1, 2, 3 and 4).
    Dependence on $\mu\equiv m\Sigma V$ is shown.
  }
  \label{fig:moments}
\end{figure}

In order to quantify the shape of the distributions, we
consider $n$-th moments
\begin{equation}
  \label{eq:moments}
  \langle \zeta_k^n \rangle = \int d \zeta_k\,
  \zeta_k^n p_k(\zeta_k; \mu),
\end{equation}
which can be calculated numerically.  The results for
$\langle(\zeta_k-\langle\zeta_k\rangle)^n\rangle$ are shown in
Figure~\ref{fig:moments} as a function of $\mu$.  From the plot for
$\langle\zeta_k\rangle$ one can see that the lowest eigenvalue is
lifted near the massless limit due to a repulsive force by the
dynamical fermions.  When $\mu$ is greater than 10, the eigenvalues
qualitatively behave as in the quenched theory (or $\mu\to \infty$
limit).  Transition from the massless two-flavor theory to the
quenched theory occurs around $\mu=$ 1--10, where the moments of the
lowest-lying eigenvalue show rather peculiar dependence on $\mu$.

The ChRMT spectrum is expected to match with those of the QCD
Dirac operator up to a constant $\Sigma V$. 
For example, the lowest eigenvalue of the QCD Dirac operator
$\lambda_1$ is matched as
\begin{equation}
  \label{eq:1steigen}
  \langle \lambda_1 \rangle / m =  
  \langle \zeta_1 \rangle / N\hat{m} = 
  \langle \zeta_1 \rangle / m\Sigma V,
\end{equation}
from which one can extract $\Sigma$, one of the fundamental
constant in ChPT.
Unlike the standard lattice QCD calculation, we do not need
any chiral extrapolation, as $m$ is already very small in
the $\epsilon$-regime.
By investigating the consistency with the determination
through higher eigenvalues or their shapes, 
one can estimate possible
systematic errors due to higher order effects in the
$\epsilon$ expansion.

\section{Numerical Simulation}
\label{sec:simulation}

\subsection{Overlap fermion implementation}
We employ Neuberger's overlap fermion formulation 
\cite{Neuberger:1997fp,Neuberger:1998wv}
for the sea quarks. 
Its Dirac operator is defined as
\begin{equation}
  \label{eq:ov}
  D(m) = 
  \left(m_0+\frac{m}{2}\right)+
  \left(m_0-\frac{m}{2}\right)
  \gamma_5 \mbox{sgn}[H_W(-m_0)],
\end{equation}
where $H_W=\gamma_5D_W(-m_0)$ denotes the Hermitian
Wilson-Dirac operator with a large negative mass $-m_0$.
We choose $m_0=1.6$ throughout this work.
(Here and in the following the parameters are given in the
lattice unit.)
The overlap-Dirac operator (\ref{eq:ov}) satisfies 
the Ginsparg-Wilson relation \cite{Ginsparg:1981bj}
\begin{equation}
  \label{eq:GW}
  D(0)\gamma_5 + \gamma_5D(0)=\frac{1}{m_0}D(0)\gamma_5D(0),
\end{equation}
when the quark mass $m$ vanishes.
Because of this relation, the fermion action built up with
(\ref{eq:ov}) has an exact chiral symmetry under the
modified chiral transformation \cite{Luscher:1998pq}.

In the practical application of the overlap-Dirac operator
(\ref{eq:ov}), the profile of near-zero modes of the kernel 
operator $H_W(-m_0)$ is important, as they determine the
numerical cost of the overlap fermion.
The presence of such near-zero modes is also a problem for
the locality property of the overlap operator 
\cite{Hernandez:1998et}.
For most gauge actions used in practical simulations, it is
known that the spectral density $\rho_W(\lambda_W)$ of the
operator $H_W(-m_0)$ is non-zero at vanishing eigenvalue
$\lambda_W$ = 0 \cite{Edwards:1998sh} due to the so-called
dislocations, 
{\it i.e.} local lumps of the gauge configuration
\cite{Berruto:2000fx}. 
We avoid this problem by introducing additional fermions and
ghosts to generate a weight
\begin{equation}
  \label{eq:detHw}
  \frac{\det[H_W(-m_0)^2]}{\det[H_W(-m_0)^2+m_t^2]},
\end{equation}
in the partition function \cite{Fukaya:2006vs}.
(The same idea is proposed in the context of the
domain-wall fermion \cite{Izubuchi:2002pq, Vranas:2006zk}.)
They are unphysical as their mass is of order of lattice
cutoff, and thus does not affect low-energy physics.
The numerator suppresses the near-zero modes, while the
denominator cancels unwanted effects for higher modes.
The ``twisted-mass'' parameter $m_t$ determines the value of
threshold below which the eigenmodes are suppressed.
We set $m_t$ = 0.2 in this work.
With these extra degrees of freedom, the spectral density
$\rho_W(\lambda_W)$ vanishes at the vanishing eigenvalue
$\lambda_W$, and the numerical cost of approximating the
sign function in (\ref{eq:ov}) is substantially reduced
\cite{Fukaya:2006vs}. 

We approximate the sign function using a rational function
of the form 
(see, {e.g.}, \cite{vandenEshof:2002ms,Chiu:2002eh}) 
\begin{equation}
  \label{eq:rational}
  \frac{1}{\sqrt{H_W^2}} = \frac{d_0}{\lambda_{min}}
  (h_W^2+c_{2n}) \sum_{l=1}^n \frac{b_l}{h_W^2+c_{2l-1}},
\end{equation}
where $\lambda_{min}$ is the lower limit of the range of 
approximation and $h_W\equiv H_W/\lambda_{min}$.
The coefficients $b_l$, $c_l$ and $d_0$ can be 
     determined analytically (the Zolotarev approximation)
     so as to optimize the accuracy of the approximation.
Since we have to fix the lower limit $\lambda_{min}$, we
calculate a few lowest-lying eigenvalues and project them
out before applying (\ref{eq:rational}) when their absolute 
value is smaller than $\lambda_{min}$.
The value of $\lambda_{min}$ is 0.144 in our simulations.
The accuracy of the approximation improves exponentially 
as the number of poles $n$ increases.
With $n=10$, the sign function $\mbox{sgn}[H_W(-m_0)]$ is
approximated to a $10^{-8}$-$10^{-7}$
level.
Since the multi-shift conjugate gradient method can be used
to invert all the $(h_W^2+c_{2l-1})^{-1}$ terms at once, the
numerical cost depends on $n$ only weakly.

In the $\epsilon$-regime the partition function and other
physical quantities show striking dependence on the global
topological charge of gauge field.
With the lattice action including (\ref{eq:detHw}) 
the topological charge never changes
during the Hybrid Monte Carlo
(HMC) simulations, which consists of molecular dynamics (MD) 
evolution of gauge field configuration. 
This is because the topology change must accompany a zero
crossing of the eigenvalue of $H_W(-m_0)$, which is
forbidden by the factor (\ref{eq:detHw}).
The gauge configuration in a fixed topological sector can
therefore be effectively sampled.
In this work the simulations are restricted in the trivial
topological sector $Q=0$ except for one quark mass parameter
for which we carry out independent simulations at $Q=-2$ and $-4$.

Here, we assume that the ergodicity of the simulation
in a fixed topological sector is satisfied even with the
determinant (\ref{eq:detHw}).
In order to confirm this, we are studying  
the fluctuation of the local topological charge density,
which will be reported in a separate paper.

\subsection{HMC simulations}
We perform two-flavor QCD simulations using the overlap
fermion for the sea quarks,
with the approximated sign function (\ref{eq:rational}) 
with $n=10$.
Lattice size is $16^3 \times 32$ throughout this work.
For the gauge part of the action, we use the Iwasaki action
\cite{Iwasaki:1985we,Iwasaki:1984cj} at $\beta$ = 2.30 and
2.35, which correspond to the lattice spacing $a$ = 0.12~fm
and 0.11~fm, respectively, when used with the extra Wilson
fermions and ghosts.
The simulation parameters are listed in
Tables~\ref{tab:para_b2.30} and \ref{tab:para_b2.35} for
$\beta$ = 2.30 and 2.35, respectively.

\begin{table}[tbp]
  \centering
  \begin{tabular}{|c|c|c|c|}
    \hline
    $m$ & traj. & $Q$ & $a$ [fm] \\
    \hline\hline
    0.015 & 10,000 & 0 & 0.1194(15) \\
    0.025 & 10,000 & 0 & 0.1206(18) \\
    0.035 & 10,000 & 0 & 0.1215(15) \\
    0.050 & 10,000 & 0 & 0.1236(14) \\
    0.050 &  5,000 & $-2$ & \\
    0.050 &  5,000 & $-4$ & \\
    0.070 & 10,000 & 0 & 0.1251(13) \\
    0.100 & 10,000 & 0 & 0.1272(12) \\
    \hline
  \end{tabular}
  \caption{Simulation parameters at $\beta$ = 2.30.}
  \label{tab:para_b2.30}
\end{table}

The configurations from the runs at $\beta$ = 2.30
are for various
physics measurements including hadron spectrum, decay
constants, form factors, bag parameters, and so on.
In this work we use them to analyze the eigenvalue spectrum.
The simulation details will be described in a separate paper
\cite{Kaneko_HMCnf2}, but we reproduce some basic parameters
in Table~\ref{tab:para_b2.30}.
They include the sea quark mass $m$, trajectory length
(the unit trajectory length is 0.5 MD time), 
topological charge $Q$ and lattice spacing $a$
determined from the Sommer scale $r_0$ (= 0.49~fm)
\cite{Sommer:1993ce} of the heavy quark potential.  
In the massless limit, the lattice spacing is found to be
0.1184(12)~fm by a linear extrapolation in $m$.
The sea quark mass at $\beta$ = 2.30 covers the region from
$m_s/6$ to $m_s$ with $m_s$ the physical strange quark
mass. 

\begin{table}[tbp]
  \centering
  \begin{tabular}{|c|c|c|c|c|c|c|c|c|c|}
    \hline
    $m$ & traj. & $m'$ &
    $\delta_{PF2}$ & $\delta_{PF1}/\delta_{PF2}$ &
    $\delta_G/\delta_{PF1}$ &
    $\langle\Delta H\rangle$ & $P_{acc}$ & 
    $\langle P\rangle$ & $a$ [fm] \\
    \hline\hline
    0.002 &  3,690 & 0.2 & 0.0714 & 1/4 & 1/5 & 0.90(23) & 0.756 & 0.62482(1) & 0.1111(24) \\
          &  1,010 & 0.2 & 0.0625 & 1/4 & 1/5 & 1.24(50) & 0.796 & 0.62479(2) & \\
    \hline
    0.020 &  1,200 & 0.2 & 0.0714 & 1/4 & 1/5 & 0.035(09) & 0.902 & 0.62480(1) & 0.1074(30) \\
    0.030 &  1,200 & 0.4 &0.0714 & 1/4 & 1/5 & 0.253(20) & 0.743 & 0.62480(2) & 0.1127(23) \\
    0.045 &  1,200 & 0.4 & 0.0833 & 1/5 & 1/6 & 0.189(18) & 0.768 & 0.62476(2) & 0.1139(29) \\
    0.065 &  1,200 & 0.4 & 0.1 & 1/5 & 1/6 & 0.098(12) & 0.838 & 0.62474(2) & 0.1175(26) \\
    0.090 &  1,200 & 0.4 & 0.1 & 1/5 & 1/6 & 0.074(19) & 0.855 & 0.62472(2) & 0.1161(24) \\
    0.110 &  1,200 & 0.4 & 0.1 & 1/5 & 1/6 & 0.052(10) & 0.868 & 0.62471(2) & 0.1182(22) \\
    \hline
  \end{tabular}
  \caption{Simulation parameters at $\beta$ = 2.35.}
  \label{tab:para_b2.35}
\end{table}

The runs at $\beta$ = 2.35 were originally intended for a
basic parameter search and therefore the trajectory length
for each sea quark mass is limited (1,200 HMC trajectories).
It is at this $\beta$ value that we performed a run in the
$\epsilon$-regime by pushing the sea quark mass 
very close to the chiral limit $m=0.002$, which is one order
of magnitude smaller than the sea quark mass in other runs.
In Table~\ref{tab:para_b2.35} we summarize several
simulation parameters.
Among them, the basic parameters are the sea quark mass $m$,
trajectory length, plaquette expectation value 
$\langle P\rangle$, and lattice spacing.
The massless limit of the lattice spacing is evaluated to be
0.1091(23)~fm using a linear extrapolation with data above
$m$ = 0.020.
This value is consistent with the result of the
$\epsilon$-regime run at $m$ = 0.002.
The other parameters are explained below.

The HMC simulation with the overlap fermion was first
attempted by Fodor, Katz and Szabo \cite{Fodor:2003bh} and
soon followed by two other groups
\cite{DeGrand:2004nq,Cundy:2005pi}.
They introduced the so-called reflection-refraction trick in
order to treat the discreteness of the HMC Hamiltonian at
the topological boundary.
This leads to a significant additional cost for dynamical
overlap fermions compared to other (chirally non-symmetric)
fermion formulations.
We avoid such extra costs by introducing the extra Wilson
fermion determinants (\ref{eq:detHw}), with which the
MD evolution never reaches the topological boundary.

In the implementation of the HMC algorithm, we introduce the 
Hasenbusch's mass preconditioner \cite{Hasenbusch:2001ne}
together with the multiple time step technique
\cite{Sexton:1992nu}. 
Namely, we rewrite the fermion determinant as 
\begin{equation}
  \label{eq:ovdet}
  \det[D(m)]^2 = \det[D(m')]^2 
  \det\left[\frac{D(m)^2}{D(m')^2}\right]
\end{equation}
by introducing a heavier overlap fermion with mass $m'$.
We then introduce a pseudo-fermion field for each
determinant. 
In the right hand side of (\ref{eq:ovdet}) the second term
is most costly as it requires an inversion of the overlap
operator with a small mass $m$.
On the other hand, the contribution to the MD force from that term can be made small by tuning
$m'$ close to $m$.
With the multiple time step technique, such small
contribution does not have to be calculated frequently,
while the force from the first term must be calculated more
often. 
We introduce three time steps:
(i) $\delta\tau_{PF2}$ for the ratio $\det[D(m)^2/D(m')^2]$, 
(ii) $\delta\tau_{PF1}$ for the preconditioner
$\det[D(m')]^2$, and
(iii) $\delta\tau_G$ for the gauge action and the extra Wilson
fermions (\ref{eq:detHw}).
By investigating the size of MD forces from each term, we
determine the time steps and the preconditioner mass $m'$ as
listed in Table~\ref{tab:para_b2.35}. 
For the run in the $\epsilon$-regime ($\beta$ = 2.35, $m$ =
0.002) we switched $\delta\tau_{PF2}$ to a smaller value in
the middle of the run, since we encounter a trajectory which
has exceptionally large MD force from the ratio
$\det[D(m)^2/D(m')^2]$ probably due to a small eigenvalue of
$D(m)$.

An average shift of Hamiltonian during a unit trajectory
$\langle\Delta H\rangle$ determines the acceptance rate
$P_{acc}$ in the HMC algorithm.
It must be $O(1)$ or less to achieve a good acceptance
rate, which is satisfied in our runs as listed in
Table~\ref{tab:para_b2.35}.
The value at $m$ = 0.002 is larger and around 0.9--1.2.
This is due to so-called ``spikes'' phenomena, {\it i.e.}
exceptionally large values ($\sim O(10-100)$) of $\Delta H$
at some trajectories.
The spikes are potentially dangerous as they may spoil the
exactness of the HMC algorithm, but we believe that this
particular run is valid 
since we have checked that the area 
    preserving condition $\langle e^{-\Delta H}\rangle=1$ is
    satisfied within statistical errors.

For the inversion of the overlap operator we use the
relaxed conjugate gradient algorithm \cite{Cundy:2004pz}.
The trick is to relax 
the convergence condition of the inner solver
as the conjugate gradient loop proceeds.
This is allowed because the change of the solution vector
becomes smaller at the later stages of the conjugate
gradient.
The gain is about a factor of 2 compared to the conventional
conjugate gradient. 
In the middle of the simulations at $\beta$ = 2.30, we
replaced the overlap solver by the one with a
five-dimensional implementation \cite{Matsufuru:2006xr}.
This is faster by another factor of 4--5 
than the relaxed conjugate gradient method.
These details of the algorithm will be discussed in a
separate paper \cite{Kaneko_HMCnf2}.

\begin{figure}[bt]
  \centering
  \includegraphics[width=10cm,clip=true]{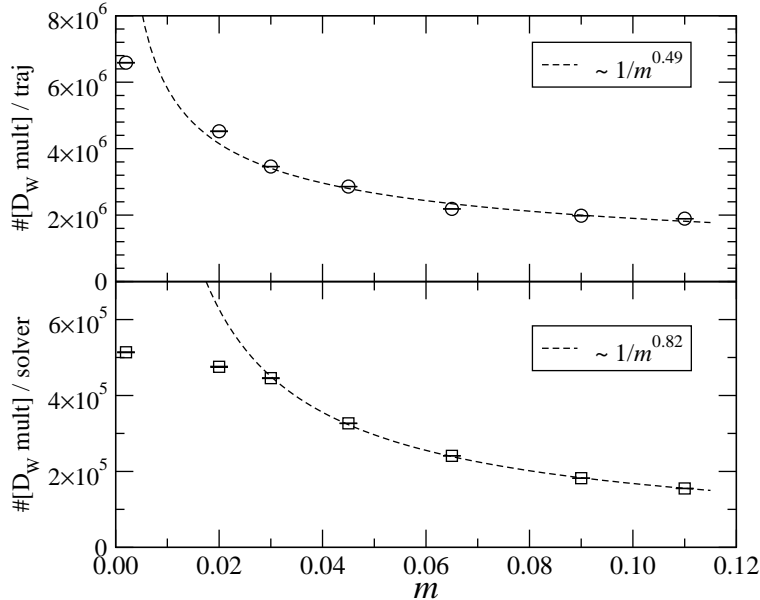}
  \caption{
    Number of the Wilson-Dirac operator multiplication per
    trajectory (upper panel) and per an overlap inversion
    (lower panel) for $\beta=2.35$.
    The curves are fits to data above $m$ = 0.030 with the
    form $\propto 1/m^\alpha$.
  }
  \label{fig:cost}
\end{figure}

The numerical cost depends on how precisely the matrix
inversions are calculated.
At an inner level there are inversions of the Hermitian
Wilson-Dirac operator appearing in the rational
approximation (\ref{eq:rational}).
The $n$ inversions can be done at the same time using the
multi-shift conjugate gradient.
We calculate until all the solutions reach the relative
precision $10^{-8}$ when adopted in the calculation of the
HMC Hamiltonian.
This value matches the precision we are aiming at for the
approximation of the sign function.
In the molecular dynamics steps the relative precision is
relaxed to $10^{-7}$.
The conjugate gradient for the overlap-Dirac operator at
the outer level is also carried out to the level of the
$10^{-8}$ ($10^{-7}$) relative precision in the HMC
Hamiltonian (MD force) calculation.

The numerical cost can be measured by counting the number of
the Wilson-Dirac operator multiplication, although other
manipulations, such as the linear algebra of vectors, are
not negligible.
The number of the Wilson-Dirac operator multiplication is
plotted in Figure~\ref{fig:cost} for the runs at $\beta=2.35$.
The upper panel shows the cost per trajectory;
the lower panel presents the cost of inverting the
overlap-Dirac operator when we calculate the Hamiltonian at
the end of each trajectory.
The expected mass dependence for the overlap solver is 
$1/\sqrt{m^2+|\lambda_1|^2}$ with $\lambda_1$ the
lowest-lying eigenvalue of the overlap operator $D(0)$.
Therefore, the cost is proportional to $1/m$ only when $m$
is much greater than $|\lambda_1|$.
This condition is satisfied for $m$ at and larger than
0.030, where $|\lambda_1|$ is around 0.004 as we show
later. 
Fitting the data with the scaling law 
$\sim 1/m^\alpha$ above $m$ = 0.030, we obtain the power
$\alpha$ as 0.82, which is roughly consistent with the
expectation.
For the total cost of the HMC Hamiltonian (upper panel), the quark mass
dependence is more significant, since it depends on the
choice of the step sizes.
It is not even a smooth function of $m$.
If we fit the data with the power law $\sim 1/m^\alpha$
above $m$ = 0.030 as in the case of the solver, we obtain 
$\alpha$ = 0.49, which gives a much milder quark mass dependence.

The machine time we spent is roughly one hour per trajectory
for the run in the $\epsilon$-regime ($m=0.002$) on a half
rack (512 computing nodes) of IBM BlueGene/L.
The cost at other mass parameters is lower as one can see in 
Figure~\ref{fig:cost}.
The numerical cost at $\beta$ = 2.30 is higher, because the
number of the near-zero modes of $H_W(-m_0)$ is
significantly larger. 

For comparison we also generated quenched configurations on
a $16^3\times 32$ lattice at $\beta$ = 2.37 in the
topological sector $Q$ = 0 and 2. 
We must use the HMC algorithm even for the quenched
simulation, as it contains the extra Wilson fermions
(\ref{eq:detHw}). 
We accumulated 20,000 trajectories for each topological
sector and used the gauge configurations for measurement at
every 200 trajectories.
The lattice spacing is 0.126(2)~fm, which matches the
dynamical lattices at $\beta=2.30$ in the heavier sea quark
mass region $m$ = 0.075 and 0.100.
In the chiral limit the dynamical lattices are slightly
finer.

\subsection{Eigenvalue calculation}
In the HMC simulations described in the previous section,
we stored the gauge configurations at every 10 trajectories 
for measurements.
For those configurations we calculate lowest 50 eigenvalues
and eigenvectors of the overlap-Dirac operator $D(0)$.
In the analysis of this work we only use the eigenvalues.

We use the implicitly restarted Lanczos algorithm for a 
chirally projected operator 
\begin{equation}
  D^+ \equiv P_+\,D(0)\,P_+,
\end{equation}
where $P_+\!=\!(1+\gamma_5)/2$. 
This operator is Hermitian and its eigenvalue gives the real
part of the eigenvalue of the original overlap operator
$D(0)$.
The pair of eigenvalues $\lambda^{ov}$ (and its complex
conjugate) of $D(0)$ can be obtained from 
${\rm Re} \lambda^{ov}$ using the relation
$|1-\lambda^{ov}/m_0|^2=1$ derived from the Ginsparg-Wilson
relation (\ref{eq:GW}).

In the calculation of the eigenvalues we enforce better
accuracy in the approximation of the sign function by
increasing the number of poles in the rational function.
The sign function is then approximated at least to the
$10^{-12}$ level.
In order to improve the convergence of the Lanczos algorithm
we use the Chebyshev acceleration technique
\cite{Neff:2001zr,DelDebbio:2005qa} and optimize the window
of eigenvalues for the target low-lying modes.

For the comparison with ChRMT, the lattice eigenvalue
$\lambda^{ov}$ is projected onto the imaginary axis as
$\lambda\equiv\mathrm{Im}\lambda^{ov}/(1-\mathrm{Re}\lambda^{ov}/(2m_0))$.
Note that $\lambda$ is very close to
$\mathrm{Im}\lambda^{ov}$ (within 0.05\%) for the low-lying 
modes we are interested in. 
We consider positive $\lambda$'s in the following.

\begin{figure}[tb]
  \centering
  \includegraphics[width=10cm,clip=true]{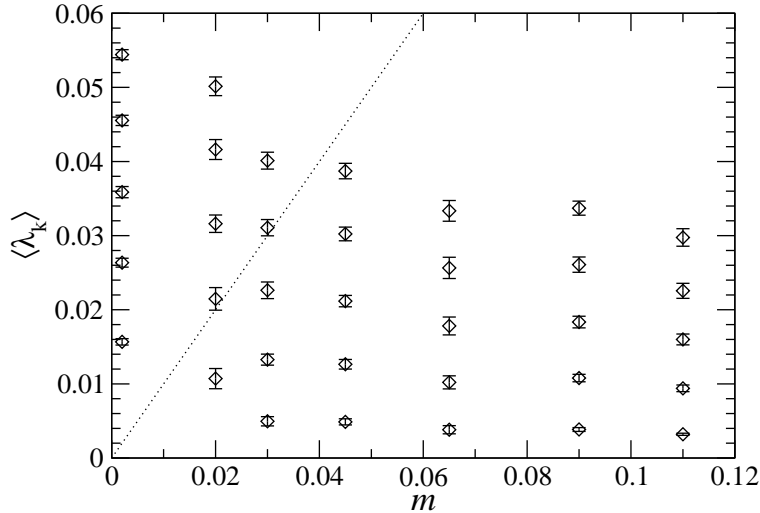}
  \caption{
    Ensemble averages of the lowest five eigenvalues
    $\langle\lambda_k\rangle$ ($k$ = 1--5) as 
    a function of sea quark mass at $\beta=2.35$.
    Dashed line shows $\lambda=m$.
  }
  \label{fig:lowest_mdep}
\end{figure}

In Figure~\ref{fig:lowest_mdep} we plot the ensemble averages
of the lowest 5 eigenvalues $\langle\lambda_k\rangle$ ($k$ =
1--5) as a function of the sea quark mass.
The data at $\beta$ = 2.35 are shown.
We observe that the low-lying spectrum is lifted
as the chiral limit is approached.
This is a direct consequence of the fermion determinant
$\sim \prod_k (|\lambda_k|^2+m^2)$, which repels the small
eigenvalues from zero when the lowest eigenvalue is larger than $m$.
This is exactly the region where the numerical cost
saturates as it is controlled by $\lambda_1$ rather than $m$.

\begin{figure}[tb]
  \centering
  \includegraphics[width=10cm,clip=true]{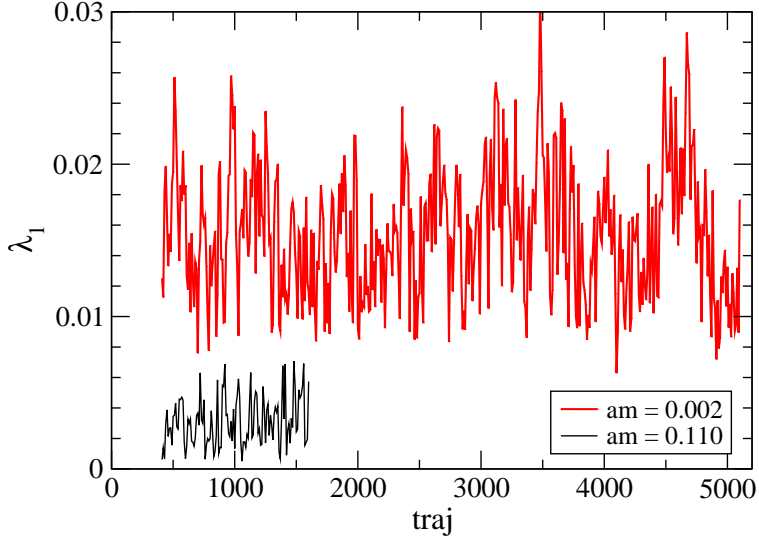}
  \caption{
    Monte Carlo history of the lowest eigenvalue $\lambda_1$
    for the sea quark masses $m$ = 0.002 (top) and 0.110
    (bottom) at $\beta$ = 2.35.
  }
  \label{fig:lowest_history}
\end{figure}

\begin{figure}[tbp]
  \centering
  \includegraphics[width=10cm]{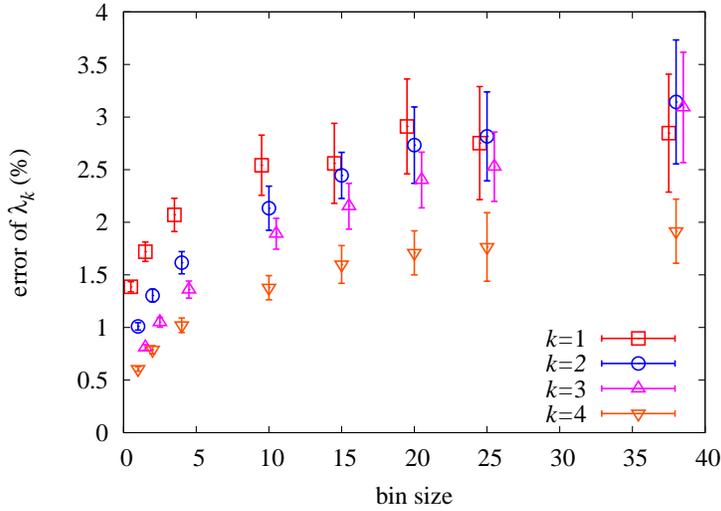}
  \caption{
    Jackknife bin-size dependence of the error for the
    eigenvalue average $\langle\lambda_k\rangle$ ($k$ =
    1--4) at $\beta=2.35$ and $m=0.002$.
    }
  \label{fig:bin-err}
\end{figure}

Figure~\ref{fig:lowest_history} shows a Monte Carlo history
of the lowest-lying eigenvalue $\lambda_1$ at the lightest
($m$ = 0.002) and the heaviest ($m$ = 0.110) sea quark
masses at $\beta$ = 2.35.
At $m$ = 0.002 we find some long range correlation extending
over a few hundred trajectories, while the history $m$ =
0.110 seems more random.
In order to quantify the effect of autocorrelation we
investigate the bin-size dependence of the jackknife error
for the average $\langle\lambda_k\rangle$ ($k$ = 1--5).
As can be seen from Figure~\ref{fig:bin-err} the jackknife
error saturates around the bin-size 20, which corresponds to
200 HMC trajectories.
This coincides with our rough estimate from
Figure~\ref{fig:lowest_history}.
In the following analysis we take the bin-size to be 20 at
$m$ = 0.002 and 10 at other sea quark masses.

\section{Low-mode spectrum in the $\epsilon$-regime}
\label{sec:low-modes}

In this section we describe a comparison of the lattice data
for the low-lying eigenvalues with the predictions of ChRMT.
The most relevant data set in our simulations is the one at
$m$ = 0.002 and $\beta$ = 2.35, since this is the only run
within the $\epsilon$-regime.

First we determine the scale, or the chiral condensate,
from the first eigenvalue through (\ref{eq:1steigen}).
By solving 
\begin{equation}
\langle\lambda_1\rangle /m =
\langle\zeta_1\rangle/\mu, \;\;\; \mu=m\Sigma V,
\end{equation}
recursively in order to correct the $\mu$ dependence of
$\langle\zeta_1\rangle$,
we obtain $\mu=0.556(16)$ and
$\Sigma^{lat} = 0.00212(6)$ in the lattice unit.
In the physical unit, the result corresponds to
$\Sigma^{lat}$ = [240(2)(6)~MeV]$^3$ 
where the second error comes from the uncertainty in
the lattice scale $a$ = 0.107(3)~fm.
In the above, we put a superscript '$\mathit{lat}$' to the chiral 
    condensate $\Sigma$ in order to emphasize that it is defined 
    on the lattice.
The error of $\langle\zeta_1\rangle=4.30$ from the 
statistical error of $\langle\lambda_1\rangle$ is neglected
(within 0.1\%). 
Note that $\mu=0.556$ is already very close to the
chiral limit as one can see from Figure~\ref{fig:moments}.
For the average of the lowest eigenvalue
$\langle\zeta_1\rangle$ the difference from the massless
limit is only 0.9\%.

\begin{figure}[tbp]
  \centering
  \includegraphics[width=10cm]{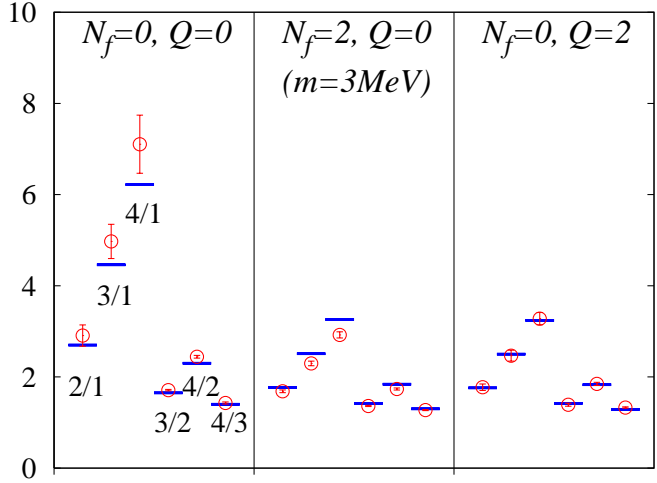}
  \caption{
    Ratio of the eigenvalues 
    $\langle\zeta_k\rangle/\langle\zeta_l\rangle$
    for combinations of $k$ and $l$ $\in$ 1--4
    (denoted in the plot as $k/l$).
 We use the input, $\mu=0.556(16)$, which is obtained
from the lowest eigenvalue average.
    In addition to the two-flavor QCD data (middle),
    quenched data at $|Q|=0$  (left) and 2 (right) at
    $\beta=2.37$ are shown. 
    Lattice data (circles) are compared with
    the ChRMT predictions (bars).
    Note that the finite $\mu (\sim 0.56)$ corrections to
    the massless case are tiny.
  }
  \label{fig:eigenratio}
\end{figure}

Next, let us compare the higher eigenvalues of the Dirac operator.
We plot the ratios 
$\langle\zeta_k\rangle/\langle\zeta_l\rangle$
of eigenvalues in Figure~\ref{fig:eigenratio}.
The lattice data agree well with the ChRMT predictions (middle panel). 
It is known that there exists the so-called
flavor-topology duality in ChRMT: the low-mode spectrum is
identical between the two-flavor (massless) theory at $Q=0$
and the quenched theory at $|Q|=2$ (right panel), while the
quenched spectrum at $Q=0$ is drastically different (left panel). 
This is nicely reproduced by the lattice data.
Note that the finite $\mu (\sim 0.56)$ corrections to 
    the massless case are very small.

\begin{figure}[tbp]
  \centering
  \includegraphics[width=10cm]{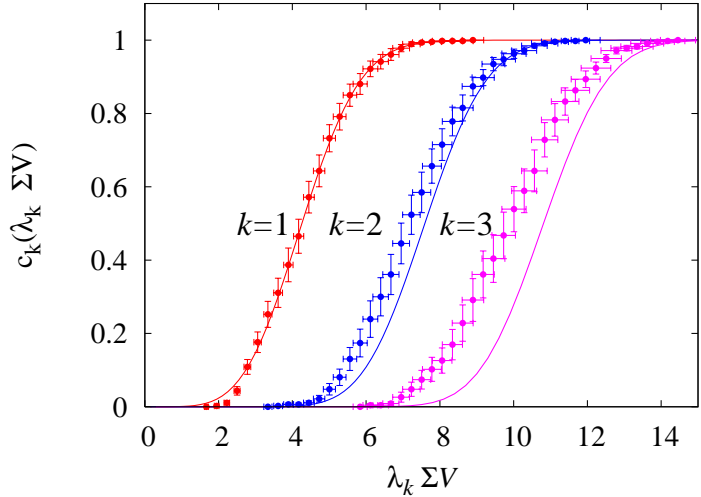}
  \caption{
    The accumulated histogram of the eigenvalues. 
    $x$-error comes from the statistical error of $\Sigma$.
    The solid lines are the ChRMT results with an input for
    $\Sigma$ from the average of the lowest eigenvalue.
  }
  \label{fig:eigenhist}
\end{figure}


\begin{table}[tbp]
  \centering
  \begin{tabular}{ccccccc}
    \hline\hline $k$
    & $\langle\zeta_k \rangle$
    & $\langle\lambda_k\rangle \Sigma V$
    & $\langle (\zeta_k -\langle\zeta_k \rangle)^2\rangle$
    & $\langle (\lambda_k-\langle\lambda_k\rangle)^2\rangle (\Sigma V)^2$
    & $\langle (\zeta_k -\langle\zeta_k \rangle)^3\rangle$
    & $\langle (\lambda_k-\langle\lambda_k\rangle)^3\rangle (\Sigma V)^3$ \\
    \hline
    1 & 4.30 & [4.30] & 1.52 & 1.48(12) & 0.41 & 0.74(27) \\
    2 & 7.62 & 7.25(13) & 1.73 & 2.11(24) & 0.28 & 0.83(43)\\
    3 & 10.83 & 9.88(21) & 1.88 & 2.52(31) & 0.22 & 0.38(58) \\
    4 & 14.01 & 12.58(28) & 2.00 & 2.39(31) & 0.18 & 0.22(66) \\
    \hline
  \end{tabular}
  \caption{
    Moments of the low-lying eigenvalues.
    Comparison between ChRMT and lattice data are made
    for the first three moments.
    The average value of the lowest eigenvalue
    $\langle\zeta_1\rangle=\langle\lambda_1\rangle\Sigma V$
    is an input for $\Sigma$.
 Here, the errors of $\langle \zeta_k\rangle$'s or their higher moments 
 due to the uncertainty of $\Sigma$
 are neglected (within 0.1\%).
  }
  \label{tab:moments}
\end{table}

Another non-trivial comparison can be made through
the shape of the eigenvalue distributions.
We plot the cumulative distribution
\begin{equation} 
  c_k(\zeta_k)\equiv\int_0^{\zeta_k}d\zeta^\prime p_k(\zeta^\prime),
\end{equation}
of the three lowest eigenvalues in Figure~\ref{fig:eigenhist}.
The agreement between the lattice data and ChRMT
(solid curves) is quite good for the lowest eigenvalue,
while
for the higher modes the agreement is marginal.
This observation can be made more quantitative by analyzing
the moments defined in (\ref{eq:moments}).
In Table~\ref{tab:moments} we list the numerical results of
both ChRMT and lattice data for the subtracted moments
$\langle(\zeta_k-\langle\zeta_k\rangle)^n\rangle$.
The overall agreement is remarkable, though we see 
deviations of about 10\% in the averages.
The deviations in the higher moments are larger in magnitude
but statistically less significant (less than two standard
deviations). 


The leading systematic error in the determination of
$\Sigma$ is the finite size effect, which scales as 
$O(\epsilon^2)\sim O(1/(F_\pi L)^2)$.
Unfortunately we can not calculate such a higher order
effect within the framework of ChRMT, but we can estimate
the size of the possible correction using the higher order
calculations of related quantities in ChPT.
To the one-loop order, the chiral condensate is written as
\begin{equation}
  \label{eq:Sigma_1loop}
  \Sigma\left[ 1 + \frac{N_f^2-1}{N_f}\frac{\beta_1}{(F_{\pi}L)^2}
  \right], 
\end{equation}
where $\beta_1$ is a numerical constant depending on the
lattice geometry \cite{Hasenfratz:1989pk}.
The value for the case of the $L^3\times (2L)$ lattice is
0.0836.
Numerically, the correction is 13\% assuming the pion decay
constant to be $F_{\pi}$ = 93~MeV.

The most direct way of reducing the systematic error is to
increase the volume, which is very costly, though.
Other possibility is to check the results with quantities
for which the higher order corrections are known.
Meson two-point functions in the $\epsilon$-regime are
examples of such quantities.
A work is in progress to calculate the two-point functions
on our gauge ensembles.

We quote the result of $\Sigma$ in the continuum
regularization scheme, {\it i.e.} the $\overline{\mbox{MS}}$
scheme.
We have calculated the renormalization factor
$Z_S^{\overline{\mathrm{MS}}}(2\mbox{~GeV})$ using the
non-perturbative renormalization technique through the RI/MOM
scheme \cite{Martinelli:1994ty}.
Calculation is done on the $\epsilon$-regime ($m=0.002$)
lattice with several different valence quark masses.
The result is 
$Z_S^{\overline{\mathrm{MS}}}(2\mbox{~GeV})=1.14(2)$.
Details of this calculation will be presented in a separate
paper.
Including the renormalization factor, our result is
\begin{equation}
  \label{eq:result}
  \Sigma^{\overline{\mathrm{MS}}}(\mathrm{2~GeV}) = 
  [251(7)(11)\mbox{~MeV}]^3.
\end{equation}
The errors represent a combined statistical error (from
$\lambda_1$, $r_0$, and
$Z_S^{\overline{\mathrm{MS}}}(2\mbox{~GeV})$) 
and the systematic error estimated from the higher order
effects in the $\epsilon$-expansion as discussed above.
Since the calculation is done at a single lattice spacing,
the discretization error cannot be quantified reliably, but
we do not expect much larger error because our lattice
action is free from $O(a)$ discretization effects.

\section{Low-mode spectrum in the $p$-regime}
\label{sec:p-regime}

For heavier sea quarks, the $\epsilon$-expansion is
not justified and the conventional $p$-expansion should be
applied instead.
Therefore, the correspondence between the Dirac eigenvalue
spectrum and ChRMT is not obvious.
On the other hand, for heavy enough sea quarks the low-lying
eigenvalues should behave as if they are in the quenched
lattices.
Here we assume that the correspondence is valid in the
intermediate sea quark mass region too, and compare the
lattice data with the ChRMT predictions for larger 
$\mu\equiv m\Sigma V$.
Strictly speaking, the theoretical connection to ChRMT
is established only at the leading order of the $\epsilon$
expansion, which is valid when
$(M_\pi L)^2\simeq (m\Sigma V)/(F_{\pi}L)^2\ll 1$ is satisfied.

\begin{figure}[tbp]
  \centering
  \includegraphics[width=10cm]{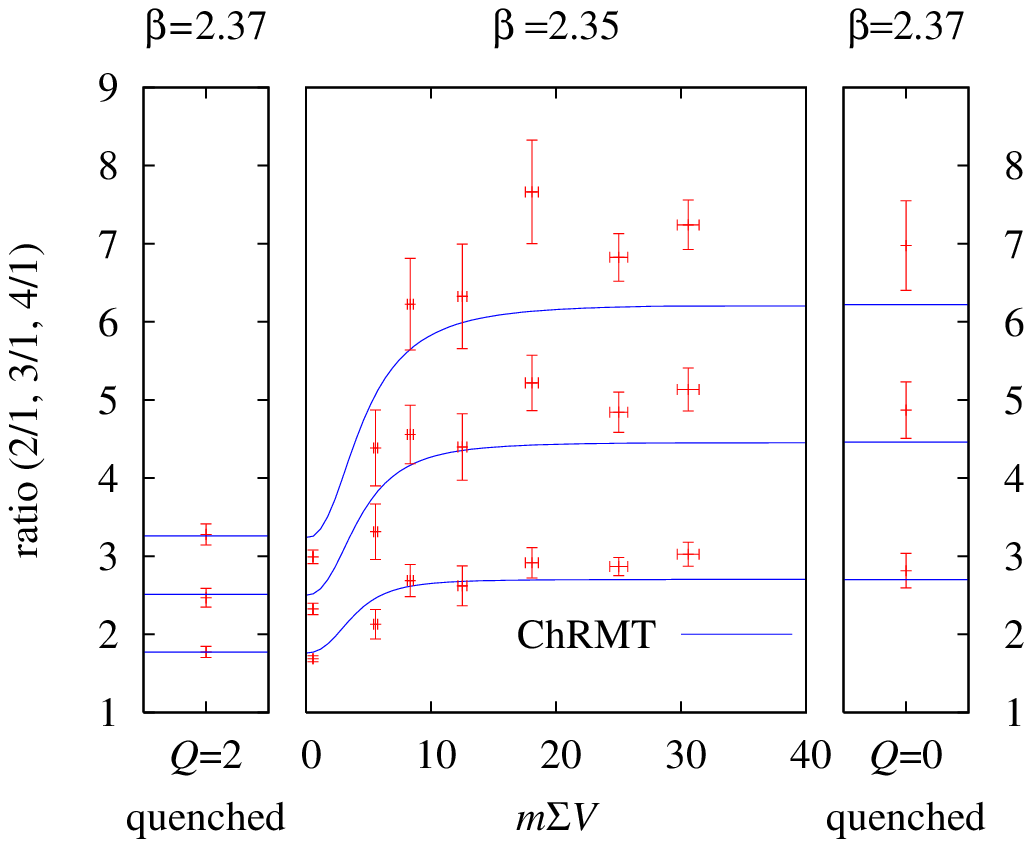}
  \includegraphics[width=10cm]{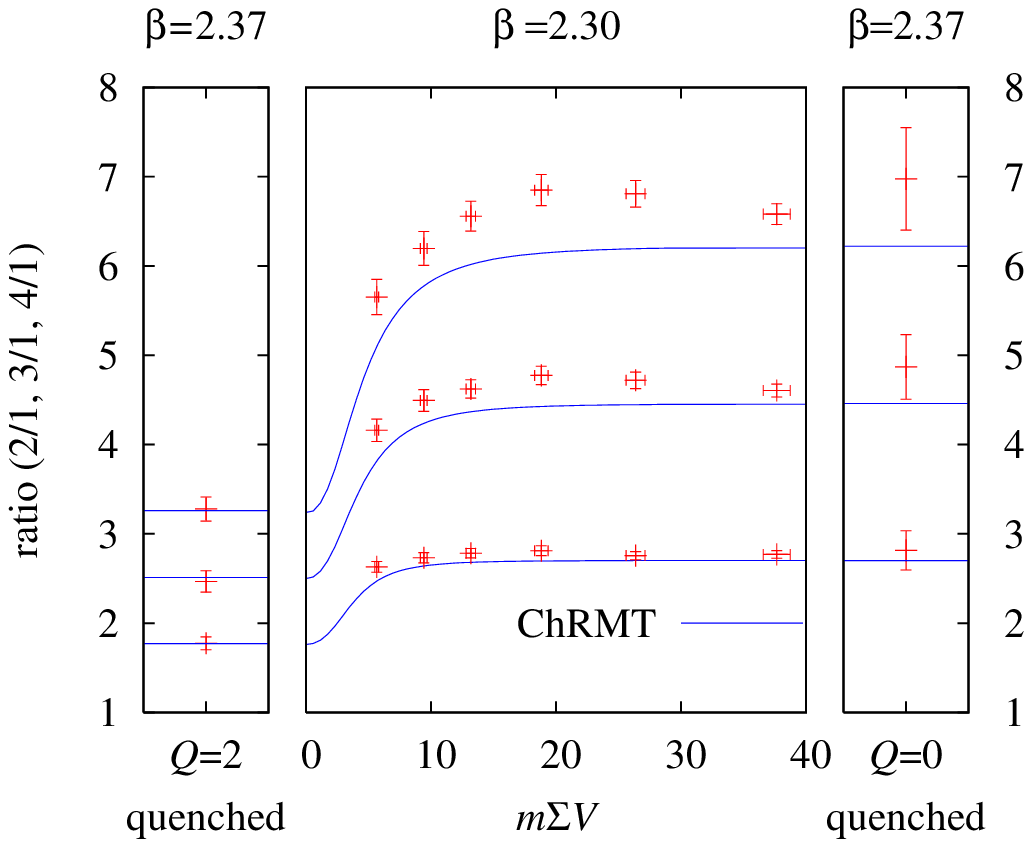}
  \caption{
    Sea quark mass dependence of the ratio of the
    eigenvalues 
    $\langle\lambda_k\rangle/\langle\lambda_1\rangle$
    for $k$ = 2, 3, and 4.
    Data at $\beta$ = 2.35 (top) and 2.30 (bottom) are
    shown. 
    Horizontal error comes from the uncertainties 
    of $\Sigma$ obtained in the $\epsilon$-regime.
    The quenched results at $\beta$ = 2.37 
    with $Q=0$ (left) and $Q=2$ (right) are also plotted to
    see the flavor-topology duality.
  }
  \label{fig:mratio}
\end{figure}

In Figure~\ref{fig:mratio} we plot the eigenvalue ratios
$\langle\lambda_k\rangle/\langle\lambda_1\rangle$
($k$ = 2--4) as a function of $m\Sigma V$.
The data are shown for both $\beta$ = 2.35 and 2.30.
The curves in the plots show the predictions of ChRMT.
The expected transition from the dynamical to quenched
lattices can be seen in the lattice data below 
$m\Sigma V\sim$ 10.
The mass dependence at $\beta$ = 2.35 is consistent with
ChRMT within relatively large statistical errors, while the
precise data at $\beta$ = 2.30 show some disagreement
especially for third and fourth eigenvalues.

\begin{figure}[tbp]
  \centering
  \includegraphics[width=10cm]{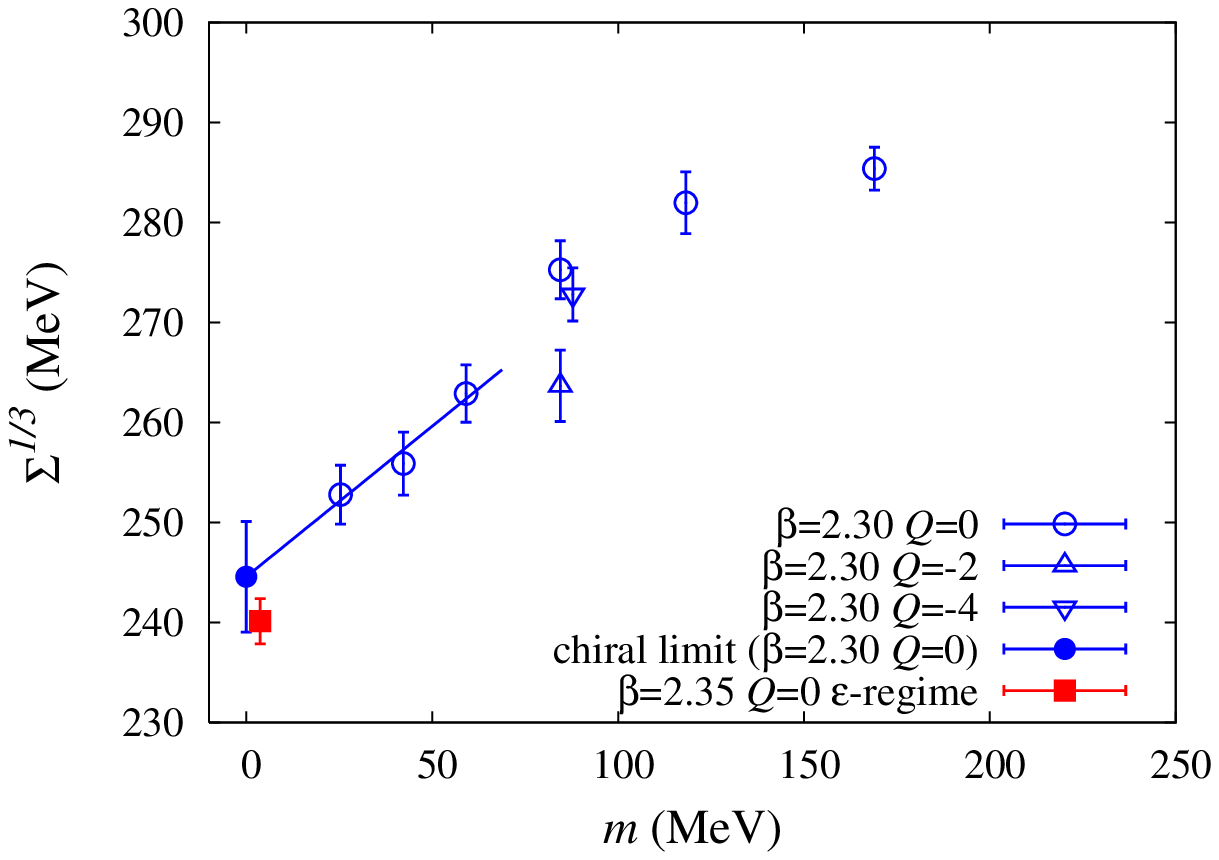}
  \caption{
    Sea quark mass dependence of the chiral condensate
    $(\Sigma^{lat})^{1/3}$
    extracted from the lowest eigenvalue.
    Open symbols denote the data at $\beta=2.30$ with
    their chiral extrapolation shown by a filled circle.
    A filled square is the result in the $\epsilon$-regime 
    ($\beta=2.35$ and $ma=0.002$).
    The lattice scale is determined through the chiral extrapolation 
    of $r_0$; its statistical error is not taken into account 
    in the plot.
  }
  \label{fig:Sigma-m}
\end{figure}

We extract the chiral condensate $\Sigma$ for each sea quark
mass using the same method applied in the $\epsilon$-regime
taking account of the mass dependence of
$\langle\zeta_1\rangle$. 
The results at $\beta$ = 2.30 are plotted in
Figure~\ref{fig:Sigma-m} (open circles).
We use a physical unit for both $m$ and $\Sigma^{lat}$;
the lattice scale is determined through $r_0$ 
after extrapolating the chiral limit.
The results show a significant sea quark mass dependence.
If we extrapolate linearly in sea quark mass using three
lowest data points we obtain
$\Sigma^{lat}$ = [245(5)(6)~MeV]$^3$
in the chiral limit.
This value is consistent with the result in the
$\epsilon$-regime as shown in the plot.

In Figure~\ref{fig:Sigma-m} we also plot data points for
non-zero topological charge ($|Q|=2$ and 4) at $m=0.050$.
We find some discrepancy between $|Q|=0$ and 2 while $|Q|=4$
is consistent with $|Q|=0$.
The size of the disagreement is about 4\% for
$(\Sigma^{lat})^{1/3}$ and thus 12\% for $\Sigma^{lat}$,
which is consistent with our estimate of the higher order
effect in the $\epsilon$ expansion.

\section{Bulk spectrum}
\label{sec:bulk}

\begin{figure}[tbp]
  \centering
  \includegraphics[width=10cm]{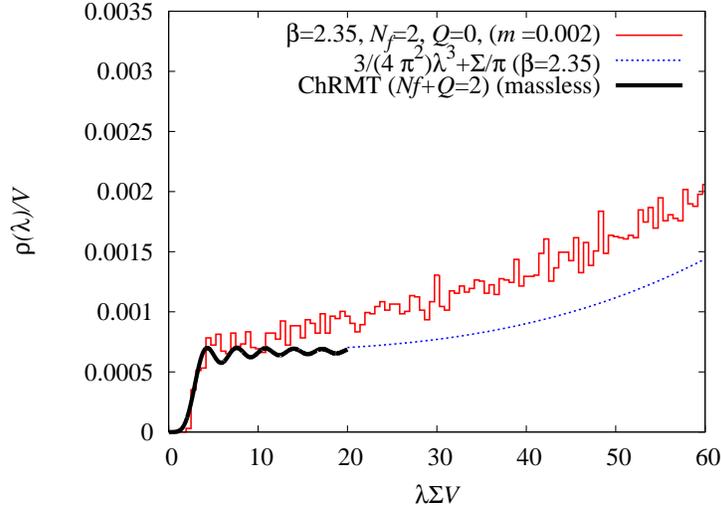}
  \caption{
    Eigenvalue histogram of the lowest 50 eigenmodes. 
 The bold curve shows the ChRMT prediction of the
 spectral density and the dashed line is 
 (free theory + constant $\Sigma/\pi$), 
 in which  we use $\Sigma=$ 0.00212 obtained in
    the $\epsilon$-regime.
  }
  \label{fig:hist}
\end{figure}

Although our data for the Dirac eigenvalue spectrum show a
qualitative agreement with the ChRMT predictions, there are
$O(10\%)$ deviations, which is significant for the larger
eigenvalues as seen in Table.~\ref{tab:moments}.
This can be understood by looking at higher eigenvalue
histogram, which we call the bulk spectrum.
Figure~\ref{fig:hist} shows a histogram of 50 lowest
eigenvalues in the $\epsilon$-regime 
($\beta$ = 2.35, $m$ = 0.002).
The normalization is fixed such that it corresponds to the
spectral density 
\begin{equation}
\rho(\lambda) \equiv \sum_k \langle \delta(\lambda-\lambda_k)\rangle,
\end{equation}
divided by the volume in
the limit of vanishing bin size.

In order to understand the shape of the data in
Figure~\ref{fig:hist} at least qualitatively, we consider a
simple model. 
Away from the low-mode region one expects a growth of the
spectral function as $\sim 3\lambda^3/4\pi^2$, which is
obtained from the number of plain-wave modes of quarks in
the free case. 
By adding the condensate contribution $\Sigma/\pi$ 
from the Banks-Casher relation \cite{Banks:1979yr}
we plot a dashed curve in
Figure~\ref{fig:hist}. 
Near the microscopic limit $\lambda\Sigma V\to 0$, the ChRMT
prediction $\Sigma\rho_{\mbox{\tiny RMT}}(\lambda\Sigma V; m\Sigma V)$
is expected to match with the data,
where $\rho_{\mbox{\tiny RMT}}$ is defined by (\ref{eq:density}).
We plot the massless case
$\Sigma\rho_{\mbox{\tiny RMT}}(\lambda\Sigma V; 0)$ 
in Figure~\ref{fig:hist} for a comparison.
(Deviation of the spectrum at $m\Sigma V=0.56$ from
the massless case is only $\sim 1$\%.).

The ChRMT curve gives a detailed description of the
Banks-Casher relation: it approaches a constant
$\Sigma/\pi$ in the large volume limit.
On the other hand, since ChRMT is valid only at the
leading order of the $\epsilon$ expansion, the region of
$O(\lambda^3)$ growth cannot be described.
Therefore, for the analysis of the microscopic eigenvalues
to be reliable, one has to work in a flat region where the
$O(\lambda^3)$ contribution is negligible.
This is the reason that the lowest eigenvalue is most
reliable to extract $\Sigma$ in our analysis in the previous
sections. 

From Figure~\ref{fig:hist} we observe that the flat region does
not extend over $\lambda\Sigma V\simeq 15$, which roughly
corresponds to the fourth lowest eigenvalue in our data.
Already at around this upper limit, the eigenvalues are
pushed from above by a repulsive force from the bulk
eigenmodes rapidly increasing as $\propto \lambda^3$, and
the ratio $\langle\lambda_k\rangle/\langle\lambda_1\rangle$
is systematically underestimated for $k$ = 3 and 4 as found
in Figure~\ref{fig:eigenratio}.
This effect is regarded as one of the finite size effect,
because the $\lambda^3$ term scales as
$(\lambda\Sigma V)^3/(\Sigma V)^3$ and its magnitude in the
microscopic regime is suppressed for larger volumes as
$1/V^3$. 
In addition, the peaks of the first few eigenvalues move
towards $\lambda\Sigma V=0$ for larger volumes, and thus
become less sensitive to the effects from bulk eigenmodes.

\begin{figure}[tbp]
  \centering
  \includegraphics[width=10cm]{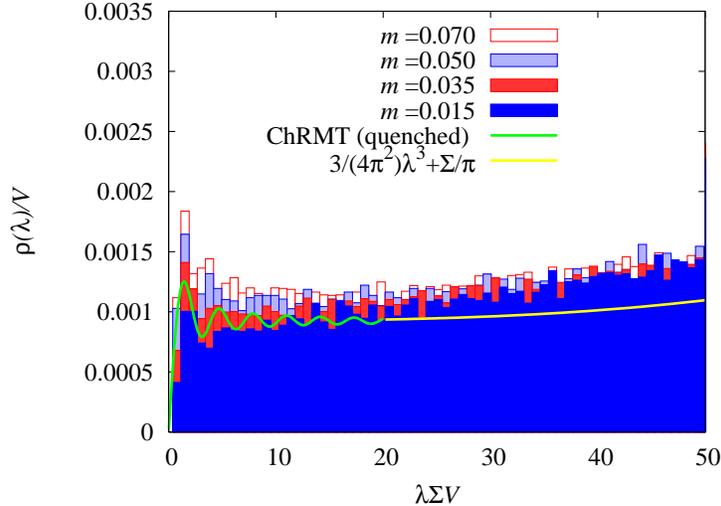}
  \caption{
    Eigenvalue histogram for the $\beta$ = 2.30 lattices.
    Solid curves show quenched ChRMT and asymptotic form
    obtained from the free quark theory.
    For a normalization we use $\Sigma=$ 0.00212 obtained in
    the $\epsilon$-regime.
  }
  \label{fig:bulkhist}
\end{figure}

The bulk spectrum for heavier sea quark masses,
which are out of the $\epsilon$-regime, 
is also interesting in order to see what happens after the
transition to the ``quenched-like'' region of the eigenvalue
spectrum.
In Figure~\ref{fig:bulkhist} the eigenvalue histogram is
shown for $\beta$ = 2.30 lattices at $m$ = 0.015, 0.035,
0.050 and 0.070, all of which are in the $p$-regime.
The plot is normalized with $\Sigma$ = 0.00212, which is
the value after the chiral extrapolation shown in
Figure~\ref{fig:Sigma-m}. 
First of all, the physical volume at $\beta$ = 2.30 is
about 30\% larger than that at $\beta$ = 2.35.
Therefore, as explained above, the growth of $O(\lambda^3)$ is expected
to be much milder and the lattice data is consistent with
this picture.
The flat region extends up to around $m\Sigma V\sim 30$.
Second, because the microscopic eigenvalue distribution
approaches that of the quenched theory, the lowest peak is
shifted towards the left. 
Overall, the number of eigenvalues in the microscopic region
increases a lot.
Unfortunately, the correspondence between ChPT and ChRMT is
theoretically less clear, since the sea quark masses are in
the $p$-regime.
In order to describe this region, the standard ChPT must be
extended to the partially quenched ChPT and a mixed
expansion has to be considered.
Namely, the sea quarks are treated in the $p$-expansion,
while the valence quarks are put in the $\epsilon$-regime to
allow the link to ChRMT.
In this paper we simply assume that ChRMT can be applied for
finite sea quark masses out of the $\epsilon$-regime.
We observe in Figure~\ref{fig:bulkhist}
that the distribution near the lowest eigenvalue is well
described by ChRMT, but the peak grows as the quark
mass increases. 
This means that the effective value of $\Sigma$ grows as the
quark mass increases, which is consistent with the sea quark
mass dependence of $\Sigma$ plotted in Figure~\ref{fig:Sigma-m}.

\section{Conclusions}
\label{sec:conclusion}

We studied the eigenvalue spectrum of the overlap-Dirac
operator on the lattices with two-flavors of dynamical
quarks.
We performed dynamical fermion simulation in the
$\epsilon$-regime by pushing the sea quark mass down to 3~MeV.
For comparison, we also calculated the eigenvalue spectrum
on the $p$-regime lattices at two lattice spacings with sea
quark mass in the range $m_s/6$--$m_s$. 
All the runs are confined in a fixed topological charge
$Q=0$, except for a few cases with finite $Q$.

We found a good agreement of the distribution of low-lying
eigenvalues in the $\epsilon$-regime with the predictions of
ChRMT, which implies a strong evidence of the spontaneous
breaking of chiral symmetry in $N_f=2$ QCD. 
We extracted the chiral condensate as 
$\Sigma^{\overline{\mathrm{MS}}}(\mathrm{2~GeV})$ = 
[251(7)(11)~MeV]$^3$ from the lowest eigenvalue.
The renormalization factor was calculated non-perturbatively.
The value of $\Sigma$ contains a systematic error of 
$\sim$ 10\% due to the higher order effect in the $\epsilon$
expansion $O(1/F_\pi L)$.
Better determination of $\Sigma$ will require larger
physical volumes to suppress such finite size effects. 

Out of the $\epsilon$-regime (the case with heavier sea
quark masses) the Dirac eigenvalue distribution still shows
a reasonable agreement with ChRMT.
The value of $\Sigma$ extracted in this region shows a
significant quark mass dependence, while its chiral limit
is consistent with the $\epsilon$-regime result.

Further information on the low-energy constants
can be extracted in the $\epsilon$-regime 
by calculating two- and three-point
functions or analyzing the Dirac eigenvalue spectrum with 
imaginary chemical potential
\cite{Damgaard:2001js, Hernandez:2006kz, Akemann:2006ru}.
The present work is a first step towards such programs.


\section*{ACKNOWLEDGMENTS}\label{sec:acknowlegments}
We thank P.H.~Damgaard and S.M.~Nishigaki for useful
suggestions and comments.
The authors acknowledge YITP workshop YITP-W-05-25 on
``Actions and Symmetries in Lattice Gauge Theory'' for
providing the opportunity to have fruitful discussions. 
Numerical simulations are performed on IBM System Blue Gene 
Solution at High Energy Accelerator Research Organization
(KEK) under a support of its Large Scale Simulation
Program (No. 07-16).
This work is supported in part by the Grant-in-Aid of the
Japanese Ministry of Education 
(No.~13135204, 15540251, 16740156, 17740171, 18340075,
18034011, 18740167, and 18840045)
and the National Science Council of Taiwan
(No.~NSC95-2112-M002-005).

\end{document}